\documentclass[traditabstract]{aa}

\usepackage{graphicx}
\usepackage{amssymb}
\usepackage{amsmath}
\usepackage{bbold}
\usepackage{url}
\usepackage{natbib}
\bibpunct{(}{)}{;}{a}{}{,}

\newcommand{\threefig}[5]{%
  \begin{figure*}%
    \centerline{\resizebox{\hsize}{!}{\includegraphics*{#1} \,%
        \includegraphics*{#2} \,%
        \includegraphics*{#3}}}%
    \caption{#5}\label{#4}%
  \end{figure*}%
}

\newcommand{\sect}[1]{Sect.~\ref{#1}}

\newcommand{\figs}[1]{Figs.~\ref{#1}}
\newcommand{\eq}[1]{Eq.~(\ref{#1})}

\renewcommand{\S}{\mathbf{S}}
\newcommand{\N}{\mathbf{N}}
\newcommand{\Nbar}{\mathbf{\bar{N}}}
\newcommand{\T}{\mathbf{T}}
\newcommand{\lmax}{\ell_{\mathrm{max}}}
\newcommand{\npix}{n_{\mathrm{pix}}}
\newcommand{\nside}{n_{\mathrm{side}}}

\newcommand{\order}[1]{${{\cal O}\! \left( #1 \right)}$}
\newcommand{\beq}{\begin{equation}}
\newcommand{\eeq}{\end{equation}}
\newcommand{\wmap}{\emph{WMAP}}
\newcommand{\planck}{\emph{Planck}}
\newenvironment{referee}{\bf}{}
\newcommand{\ber}{\begin{referee}}
\newcommand{\eer}{\end{referee}}

\begin{document}

\title{Efficient Wiener filtering without preconditioning}

\titlerunning{}

\author{Franz Elsner\inst{1}
  \and
  Benjamin D. Wandelt\inst{1,2}}

\institute{Institut d'Astrophysique de Paris, UMR 7095, CNRS -
  Universit\'e Pierre et Marie Curie (Univ Paris 06), 98 bis blvd
  Arago, 75014 Paris, France\\
  \email{elsner@iap.fr}
  \and
  Departments of Physics and Astronomy, University of Illinois at
  Urbana-Champaign, Urbana, IL 61801, USA}

\date{Received \dots / Accepted \dots}

\abstract{We present a new approach to calculate the Wiener filter
  solution of general data sets. It is trivial to implement, flexible,
  numerically absolutely stable, and guaranteed to converge. Most
  importantly, it does not require an ingenious choice of
  preconditioner to work well. The method is capable of taking into
  account inhomogeneous noise distributions and arbitrary mask
  geometries. It iteratively builds up the signal reconstruction by
  means of a messenger field, introduced to mediate between the
  different preferred bases in which signal and noise properties can
  be specified most conveniently. Using cosmic microwave background
  (CMB) radiation data as a showcase, we demonstrate the capabilities
  of our scheme by computing Wiener filtered \wmap7 temperature and
  polarization maps at full resolution for the first time. We show how
  the algorithm can be modified to synthesize fluctuation maps, which,
  combined with the Wiener filter solution, result in unbiased
  constrained signal realizations, consistent with the
  observations. The algorithm performs well even on simulated CMB maps
  with \planck\ resolution and dynamic range.}


\maketitle

\section{Introduction}
\label{sec:intro}

Signal reconstruction out of noisy data is arguably one of the most
fundamental problems in astronomy (and science in general), and has
been subject of extensive research for centuries. The method of least
squares, for example, was originally introduced to predict the orbital
parameters of the dwarf planet 1 Ceres out of a series of astrometric
measurements \citep{1809gauss}. Since then, applications in signal
processing have become legion.

Incorporating statistical information about the properties of noise
and signal laid the foundation for a large class of powerful
reconstruction techniques, including particularly popular methods such
as the Wiener filter \citep{1949wiener}, approaches that rely on
entropy maximization \citep{1957PhRv..106..620J}, or the Kalman filter
\citep{1960kalman}. The issue has remained a topic of ongoing
research. For Gaussian fields, only recently efficient algorithms have
been developed for cases without prior knowledge of the signal
properties \citep[e.g.,][]{2004PhRvD..70h3511W, 2010MNRAS.406...60J,
  2011PhRvD..83j5014E}, or, where neither signal nor noise covariances
are known a priori \citep{2011PhRvE..84d1118O}.

In this letter, we focus on the evaluation of the generalized Wiener
filter. If the observed data $d$ are a linear combination of signal
$s$ and noise $n$,
\beq
\label{eq:def_data}
d = s + n\, ,
\eeq
the Wiener filter is defined as the solution of the equation
\beq
\label{eq:def_wiener}
(\S^{-1} + \N^{-1}) \, s_{\mathrm{WF}} = \N^{-1} d \, .
\eeq
Then, by construction, $s_{\mathrm{WF}}$ is the maximum a posteriori
probability solution if signal and noise are both Gaussian random
fields with known covariances $\S$ and $\N$,
respectively. 

The Wiener filter or variants thereof are frequently encountered in
the post-processing of observational data. In the analysis of cosmic
microwave background (CMB) radiation experiments, for example, during
optimal power spectrum estimation \citep[e.g.,][]{1997PhRvD..55.5895T,
  1998PhRvD..57.2117B, 1999ApJ...510..551O, 2012A&A...540L...6E},
likelihood analysis \citep[e.g.,][]{2007ApJS..170..288H,
  2009ApJS..180..306D, 2012A&A...542A..60E}, mapmaking
\citep[e.g.,][]{1997ApJ...474L..77T, 1997ApJ...480L..87T}, and lensing
reconstructions \citep[e.g.,][]{2004PhRvD..70j3501H,
  2007PhRvD..76d3510S}, etc.

In practice, the numerical computation of \eq{eq:def_wiener} can be
challenging for the large data sets collected by modern
experiments. The underlying difficulty is that signal and noise
covariances grow as the square of the size of the data set and become
too large to be stored and processed as dense matrices. Progress is
possible when these matrices are structured, e.g. when there are sets
of bases where $\S$ and $\N$ become sparse. However, the bases may be
incompatible, for example, the signal covariance may be sparse in
Fourier space, whereas the noise covariance may be sparse in pixel
space. As a result, one can either solve the Wiener filter equation
only approximately, e.g. by assuming a homogeneous and isotropic noise
distribution \citep[as done in, e.g.,][]{2004PhRvD..70j3501H,
  2005ApJ...634...14K, 2009PhRvD..80l3007M}, or has to rely on complex
and costly numerical schemes \citep[e.g., involving conjugate gradient
  solvers with multigrid
  preconditioners,][]{2007PhRvD..76d3510S}. Finding fast
preconditioners that work is a highly non-trivial art especially since
the matrices of interest are often extremely ill-conditioned, with
typical condition numbers of the order \order{10^7}.

The Wiener filter is the optimal linear filter only for strictly
Gaussian noise and signal. However, a maximum entropy argument shows
that the Gaussian prior is the least informative prior, once mean and
covariance of the random fields are specified. From a Bayesian point
of view, a Gaussian prior can therefore be a well-justified
approximation even for non-Gaussian inference. This argument provides
a theoretical underpinning for the work of, e.g,
\citet{1999MNRAS.302..663B}, where the authors use Wiener filtering as
a means to clean CMB maps from (non-Gaussian) foregrounds. Applied in
the limit of weak non-Gaussianity, the Wiener filter has also proven
an indispensable tool for the search for primordial non-Gaussianity in
CMB data \citep{2005ApJ...634...14K}, where inference from the full
non-Gaussian posterior is computationally very demanding
\citep{2010ApJ...724.1262E}.

In this letter, we propose a high precision, iterative algorithm for
the solution of the exact Wiener filter equation. It is conceptually
very simple and therefore easy to implement correctly, yet still
numerically efficient enough to be of general interest. The method is
fully universal as long as there exist easily accessible orthogonal
basis sets which render $\S$ and $\N$ sparse. We illustrate our
approach by applying it to CMB signal reconstruction.

The letter is organized as follows. In \sect{sec:method}, we introduce
our new algorithm for the efficient calculation of the Wiener
filter. We then present an explicit example and analyze CMB data of
the \wmap\ experiment (\sect{sec:wmap}) showing how to generalize
\eq{eq:def_data} to include an instrumental transfer
function. Finally, we summarize our findings in
\sect{sec:summary}. Throughout the paper, we assume WMAP7+BAO+H0
cosmological parameters \citep{2011ApJS..192...18K}.

\section{Method}
\label{sec:method}

In the following, we present the basic equations of the algorithm and
address how to solve them numerically before discussing a concrete
example.

\subsection{Algorithm for Wiener filtering}
\label{subsec:wiener_filter}

For an efficient evaluation of \eq{eq:def_wiener}, we introduce an
additional stochastic auxiliary field $t$ with covariance $\T$. The
statistical properties of this helper variable are deliberately chosen
very simple: $\T$ is proportional to the identity matrix. The identity
matrix is invariant under \emph{any} orthogonal basis change. While it
may not be possible to directly apply expressions like $(\S +
\N)^{-1}$ to a data vector, we can always do so with combinations as
$(\S + \T)^{-1}$ and $(\N + \T)^{-1}$
regardless of which basis renders $\S$ or $\N$ sparse.

In order to take advantage of this possibility, we form a set of
equations to be solved iteratively for the signal reconstruction $s$
and the auxiliary field $t$,
\begin{align}
 \label{eq:def_algorithm1}
    \left( \Nbar^{-1} + (\lambda \T)^{-1} \right) \,
    t & = \Nbar^{-1} \, d + (\lambda \T)^{-1} \, s
    \\
 \label{eq:def_algorithm2}
    \left( \S^{-1} + (\lambda \T)^{-1} \right)
    \, s & = (\lambda \T)^{-1} \, t \, ,
\end{align}
where we defined $\Nbar \equiv \N - \T$ and introduced one additional
scalar parameter $\lambda$, whose usefulness we will discuss in the
context of convergence acceleration in \sect{subsec:convergence}. We
choose the amplitude of the covariance matrix of the auxiliary field
according to $\mathbf{T} = \tau \mathbb{1} $ with
$\tau\equiv\min(\mathrm{diag}(\mathbf{N}))$. Consequently, the field
$t$ gains a physical interpretation: it corresponds to a signal
reconstruction given the fraction of the noise that is distributed
homogeneously and isotropically in the data. It can be easily shown
that the above system reduces to the Wiener filter equation in the
limit $\lambda = 1$.

The algorithm can be summarized as follows. We initialize the vectors
$s$ and $t$ with zeros. First, we solve \eq{eq:def_algorithm1} for the
auxiliary field $t$ in the basis defined by the noise covariance
matrix. Next, we change the basis to, e.g., Fourier space, where $\S$
can be represented easily. Then, we solve for $s$ using
\eq{eq:def_algorithm2} given the latest realization of $t$, and
finally transform the result back to the original basis. After a
sufficient number of iterations, the signal reconstruction converges
to the Wiener filter solution, $s \rightarrow s_{\mathrm{WF}}$ as we
show now.

\subsection{Stability, convergence speed and diagnostics}
\label{subsec:convergence}

The method is unconditionally stable. If the residual is
$\epsilon_{i}$ at one iteration, it will be
\beq
\label{eq:epsilon}
\epsilon_{i+1}=\left[ \S (\S + \lambda \T)^{-1} \right] \left[
  \Nbar (\Nbar + \lambda \T)^{-1} \right] \epsilon_{i}
\eeq
at the next. If $\N \propto \mathbb{1}$, the system is solved in a
single iteration. The terms in square brackets have spectral radii
$\rho_{1, \, 2}$ strictly less than unity. So $|\epsilon_{i+1}| <
\rho_{1} \rho_{2} \, |\epsilon_{i}|$ for all $i$, and convergence is
exponential. It is fast for low noise pixels and signal modes with low
prior variance, and slowest for modes with high prior variance in high
noise pixels. Temporarily setting $\lambda> 1$ accelerates convergence
in this problematic regime.

For a given $\lambda$, the iterations minimize
\beq
\chi^2(x,\lambda) = x^{\dagger} \S^{-1} x + (d - x)^{\dagger}
\left[ \N + (\lambda - 1) \T \right]^{-1} (d - x)
\eeq
to give $s(\lambda) = \S (\S + \N + (\lambda-1)\T)^{-1}$. Fortunately,
it is precisely in the regime where convergence is slowest ($\S$, $\N$
large) that $s(\lambda>1)$ approximates $s(1)$ best.

The goal is therefore to find a ``cooling schedule'' which starts at
high $\lambda$ and reduces it to $\lambda = 1$ such that the final
solution satisfies a defined accuracy criterion. Note that further
iterations at lower $\lambda$ never degrade but always improve parts
of the solution that were found at higher $\lambda$.

At $x = s(\lambda)$, we have $\chi_\mathrm{min}^{2}(\lambda) =
d^{\dagger} (\S + \N + (\lambda-1)\T)^{-1} d = d^{\dagger} \S^{-1}
s(\lambda) < \chi_{\mathrm{min}}^{2}(1)$ for $\lambda > 1$. We now
show that we can find a ``cooling schedule'' for $\lambda$ which
respects the error bound $E = \chi^{2}(\lambda, x) < b =
\chi^{2}_\mathrm{min}(1)$ throughout.

The following algorithm has the desired property: initialize $x = 0$
and choose the initial $\lambda$ such that $\chi^{2}(\lambda, 0) =
d^{\dagger} (\N + (\lambda-1)\T)^{-1} d = b$. Then iterate $x$ once at
that $\lambda$, which reduces $E$ to $\chi^{2}(\lambda, x)$. Decrease
$\lambda$ such that $E = b$. Repeat until $\lambda = 1$ at which point
$x = s(1)$.

So far we have not specified how to calculate the optimal $b =
\chi^{2}_\mathrm{min}(1)$, which only becomes easy to compute
\emph{after} the Wiener Filter has been found. Choosing $b >
\chi^{2}_\mathrm{min}(1)$ leads to a needlessly inaccurate map,
containing numerical noise artifacts that could result in erroneous
statistical conclusions; choosing $b < \chi^{2}_\mathrm{min}(1)$,
gives results that are not wrong, just suboptimal.

Fortunately, it is possible to approximate $b$ during the solution, as
long as the initial $b < \chi^{2}_\mathrm{min}(1)$. After converging
at $\lambda > 1$ when $b < \chi^{2}_\mathrm{min}(1)$, we can increase
$b$ to make it sufficiently close to $\chi^{2}_\mathrm{min}(1)$ in one
step with minimal additional computation. At that point, we have $x =
s(\lambda)$ and can calculate $b' = \chi^{2}_\mathrm{min}(\lambda) +
(\lambda-1)\tau x^{\dagger} \S^{-2} x < \chi^{2}_\mathrm{min}(1)$,
with the inequality being quadratically small.

If $d$ is consistent with the assumed statistical model, $d = s + n$,
we can find how close $b$ should be to $\chi_\mathrm{min}^{2}(1)$. For
an ensemble of data sets, $\chi_\mathrm{min}^{2}(1)$ is
$\chi^2$-distributed with a mean given by the number of degrees of
freedom, $n_{\mathrm{d.o.f.}}$, and a variance of $\sigma^2_{\chi^2} =
2 \, n_{\mathrm{d.o.f.}}$. Hence, no outcome of statistical inference
based on the resulting Wiener Filter depends on whether it has an
excess $\chi^{2}(1)$ of much less than $\sigma_{\chi^{2}}$.

Even if we had chosen $b$ far below the expectation, $b = n_{d.o.f} -
m \sigma_{\chi^{2}}$, with $m \gtrsim 5$, excluding a violation of $b
< \chi^{2}_\mathrm{min}(1)$ for all practical purposes, the updated
$b'$ satisifies $\chi^{2}_\mathrm{min}(1) - b'\sim \sigma_{\chi^{2}} /
\sqrt{n_\mathrm{d.o.f.}}$. It is therefore statistically irrelevant as
long as $n_\mathrm{d.o.f.} \gg 1$ which is true for all cases where an
iterative scheme is likely to be useful.

This algorithm guarantees satisfying a given error bound, but it does
not guarantee the fastest possible time to solution. For applications
to the CMB, where the signal power spectrum tends to decrease with
increasing scale, it is much cheaper to iterate at high $\lambda$ than
at low $\lambda$. In \sect{sec:wmap}, we describe a heuristic
choice for the ``cooling schedule'' that exploits the assumed shape of
the signal power spectrum and leads to significantly faster execution
times.

\subsection{Constrained realizations}

The Wiener filter is known to remove noise very aggressively. As a
result, the covariance of the signal reconstruction is reduced
compared to the true signal,
\beq
\langle s_{\mathrm{WF}}^{\vphantom{\dagger}} \,
s_{\mathrm{WF}}^{\dagger} \rangle = \S \left( \S+ \N \right)^{-1} \S \, .
\eeq
To obtain signal realizations with correct covariance properties that
are consistent with the observed data, we have to add a fluctuation
vector $f$ to the Wiener filter solution with covariance
(\citealt{1987ApJ...323L.103B}, see also, e.g.,
\citealt{1991ApJ...380L...5H})
\beq
\langle f \, f^{\dagger} \rangle = \left( \S^{-1} + \N^{-1}
\right)^{-1} .
\eeq
Applying only minor modifications, the algorithm as introduced above
can be adapted to generate suitable fluctuation maps. The only
difference lies in the source term on the right-hand side of
\eq{eq:def_algorithm1}, which we replace by
\beq
\Nbar^{-1} \, d \rightarrow \Nbar^{-1} \N \left( \N^{-1/2} \, g_{1} +
\S^{-1/2} \, g_{2} \right) \, ,
\eeq
where $g_{1, 2}$ are independent univariate Gaussian random
variables. For those entries of the vector $\S^{-1/2} \, g_{2}$ that
fall within masked regions where $\Nbar^{-1}$ is zero and formally no
inverse $\N$ can be defined, the corresponding blocks of the matrix
product $\Nbar^{-1} \N$ have to be replaced with the identity matrix.

After the algorithm has been evaluated as explained in
\sect{subsec:wiener_filter}, the solution is a random realization of a
fluctuation map $f$ with the correct covariance
properties. Constrained realizations are then given by
$s_{\mathrm{CR}} = s_{\mathrm{WF}} + f$.

\section{\wmap\ polarization analysis}
\label{sec:wmap}

We now discuss the application of our technique to Wilkinson microwave
anisotropy probe (\wmap) data \citep[][]{2011ApJS..192...14J}. We used
the seven-year observations of the CMB radiation in the V-band at
61~GHz as a basis for the reconstruction\footnote{Available at
  \url{http://lambda.gsfc.nasa.gov}}, at resolution parameter $\nside
= 512$ in the Healpix representation \citep{2005ApJ...622..759G} and
$\lmax = 1024$. Since CMB anisotropies are isotropic and Gaussian, the
signal covariance is diagonal in spherical harmonic space. The noise
covariance, on the other hand, can be approximated by a diagonal
matrix in pixel space \citep{2003ApJS..148...63H}.

However, as we are interested in the Wiener filtering of temperature
and polarization data, $\S$ and $\N$ are in fact block diagonal, with
a significantly increased condition number of the matrices
\citep{2007ApJ...656..653L}. For all multipole moments $\ell$, we
write,
\beq
S_{\ell} = b_{\ell}^2
 \begin{pmatrix}
  C^\mathrm{TT}_{\ell} & C^\mathrm{TE}_{\ell}\\
  C^\mathrm{TE}_{\ell} & C^\mathrm{EE}_{\ell}
 \end{pmatrix} \, ,
\eeq
where the $C^\mathrm{XX}_{\ell}$ are the power spectra for temperature
(XX = TT), polarization (XX = EE), and their cross-spectrum (XX = TE),
and the $b_{\ell}$ are the beam function. As we did not aim at a
reconstruction of a possible contribution from a curl component to the
polarization signal (XX = BB), we excluded it from the analysis. The
blocks of the noise covariance matrix are, for every pixel,
\beq
N_i =
 \begin{pmatrix}
  \mathrm{\langle I \, I \rangle} & 0 & 0\\
  0 & \mathrm{\langle Q \, Q \rangle} & \mathrm{\langle Q \, U \rangle}\\
  0 & \mathrm{\langle Q \, U \rangle} & \mathrm{\langle U \, U \rangle}
 \end{pmatrix} \, ,
\eeq
where we used the Stokes parameters I, Q, and U. For \wmap\ data, the
cross-correlations $\mathrm{\langle I \, Q \rangle}$ and
$\mathrm{\langle I \, U \rangle}$ are small
\citep{2007ApJS..170..335P}.

We adopted the official \wmap\ extended temperature and polarization
masks, restricting the sky fraction available to 70.6~\% and 73.3~\%,
respectively. We removed possible residual monopole and dipole
contributions after the masks had been applied. Avoiding numerical
errors associated with spherical harmonic transforms on the cut sky
can be done by an explicit low rank update of the inverse noise
covariance matrix $\mathbf{\bar{N}}^{-1}$ in \eq{eq:def_algorithm1} to
assign infinite variance to a set of template maps, representing the
(non-cosmological) modes to be subtracted
\citep{2009JCAP...09..006S}. However, we found it numerically more
stable to split the calculation into separate steps; we therefore
multiplied the input independently by
\beq
d = \lim_{\sigma \rightarrow \infty} \left(\mathbb{1} + \sigma V \,
  V^{\dagger} \right)^{-1} d^{\mathrm{raw}} \, ,
\eeq
where $V$ is a $4 \times \npix$ matrix, storing the four masked
templates to be marginalized over. We used the Woodbury matrix
identity to evaluate the equation.

To initialize the algorithm, we chose $\lambda$ according to the ratio
$T / S^\mathrm{TT}_{\ell_{\mathrm{iter}}}$ at $\ell_{\mathrm{iter}} =
20$. This starting point is well suited to let all large-scale modes
from $\ell \approx 30$ down to the quadrupole converge. After the
$\chi^2$ has reached a plateau, we decreased it to the ratio at
$\ell^{\mathrm{new}}_{\mathrm{iter}} = 2 \,
\ell^{\mathrm{old}}_{\mathrm{iter}}$. We restricted the maximal
increment in $\ell_{\mathrm{iter}}$ to 500 and assured that $\lambda
\geq 1$ is always fulfilled. We achieved robust results after
terminating the algorithm once $\Delta \chi^2 < 10^{-4} \,
\sigma_{\chi^2}$ at $\lambda = 1$ has been reached, in this case at a
final $\chi^2 / n_{\mathrm{d.o.f.}} = 0.996$. The success of the
algorithm does not sensitively depend on the scheme for lowering
$\lambda$ to unity.

The gradual decrease of $\lambda$, coupled to a well specified
multipole moment $\ell_{\mathrm{iter}}$, allows for a significant
speedup of the algorithm. Since the signal is build-up from large to
small scales, fluctuations at multipoles $\ell > \ell_{\mathrm{iter}}$
will be suppressed. As a consequence, we can restrict the maximum
multipole moment of the computationally expensive spherical harmonic
transforms to $\lmax^{\mathrm{SHT}} \approx \ell_{\mathrm{iter}} +
100$, considerably smaller than $\lmax$ for most of the
iterations. Owing to a time complexity of \order{\nside^{\vphantom{2}}
  \lmax^2}, the overall computing time for the spherical harmonic
transforms can be substantially reduced. For the final iterations
including the full range of $\ell$, the algorithm is an excellent
candidate for further acceleration by means of the massive parallelism
offered by modern graphics processing units
\citep{2011A&A...532A..35E}.

The result of the \wmap\ analysis is shown in \figs{fig:wiener_temp}
and \ref{fig:wiener_pol}, where we plot patches of the Wiener filtered
maps for the Stokes parameters I, Q, and U, as well as an example of a
temperature fluctuation map. The non-vanishing cross-spectrum
$C^\mathrm{TE}_{\ell}$ allows for an efficient polarization
reconstruction in masked regions, as long as temperature data are
available. A distinguishing feature of Wiener filtering is the
extrapolation of signal into the mask; this works especially well in
the small areas where single point sources have been cut out.

\threefig{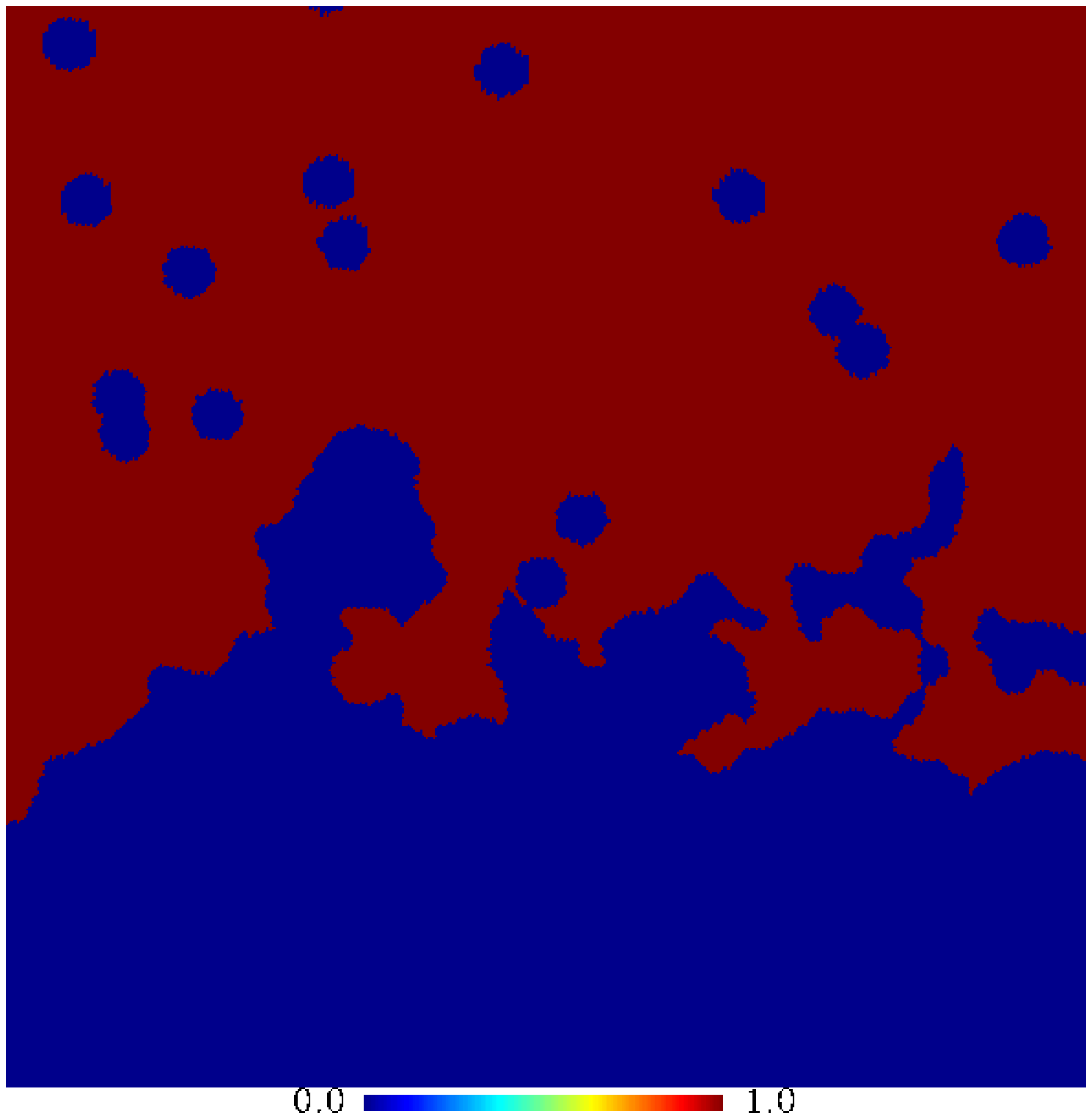}{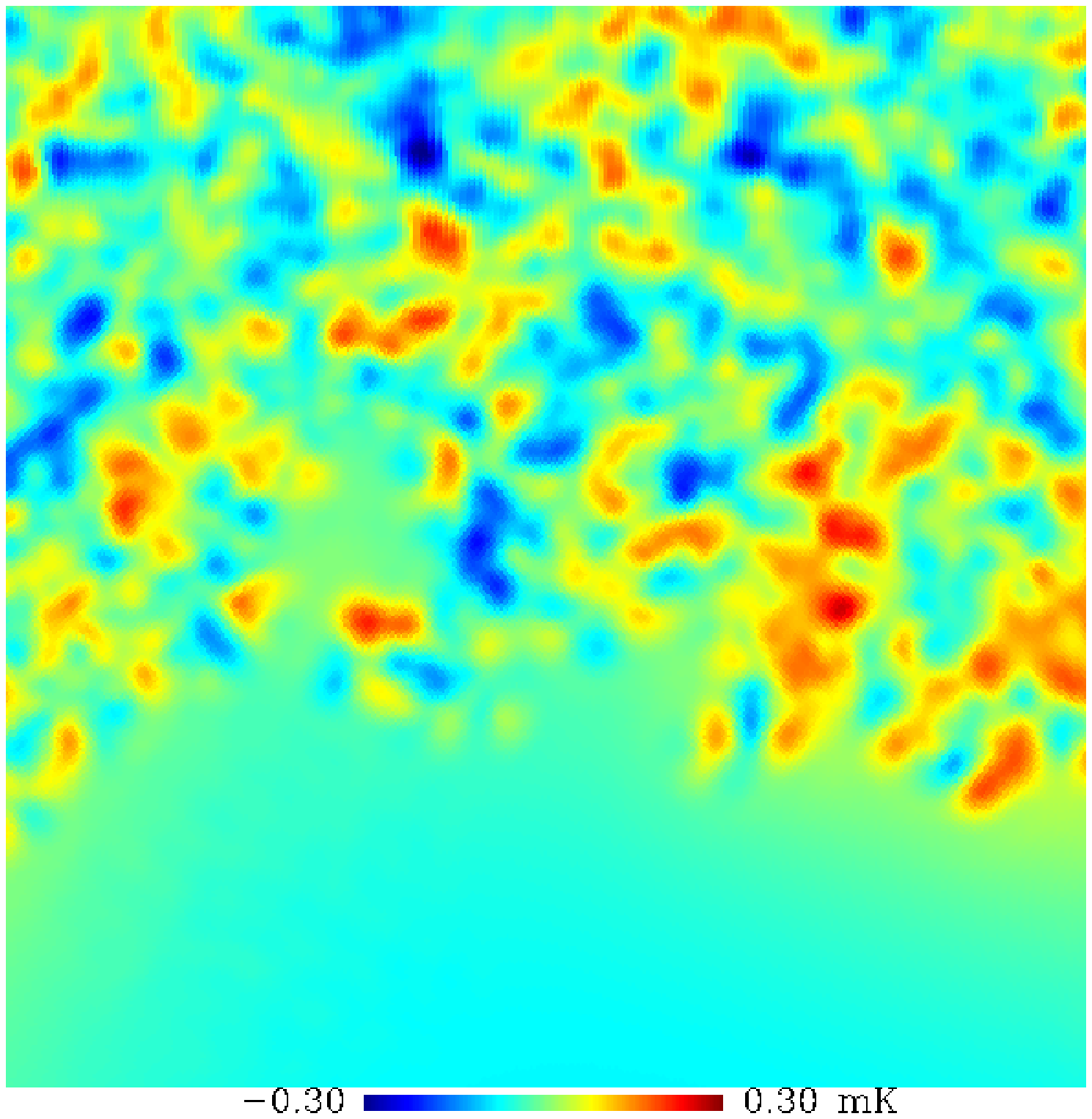}{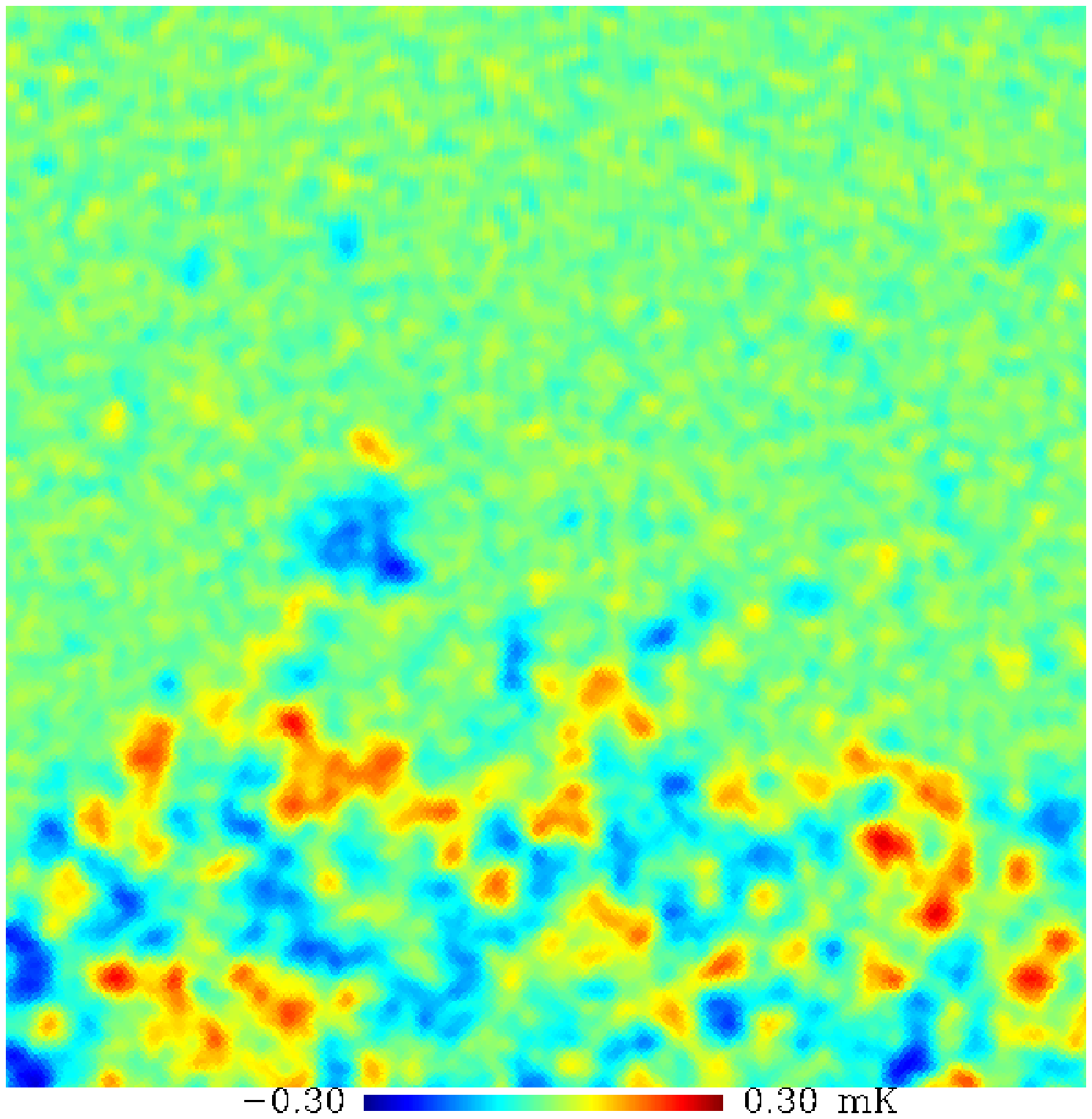}{fig:wiener_temp}%
{Temperature reconstruction. Using the \wmap\ mask (\emph{left panel})
  for the analysis, we plot the V-band temperature map after Wiener
  filtering (\emph{middle panel}). It can be supplemented to a
  constrained realization by superposing a fluctuation term
  (\emph{right panel}). The patches are $25 \degr$ on the side,
  centered at galactic coordinates $(l, \, b) = (0 \degr, \, 35
  \degr)$.}

\threefig{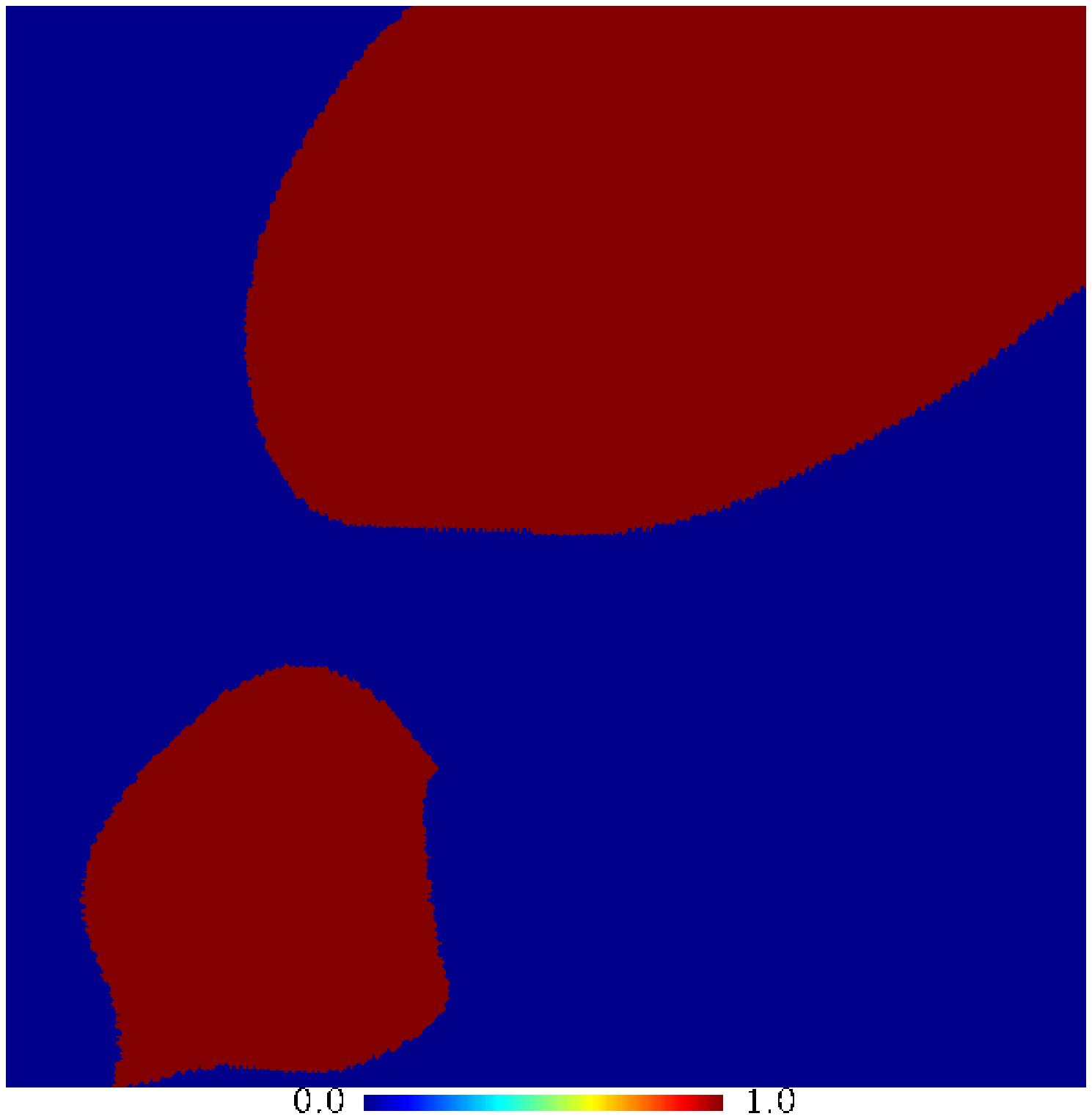}{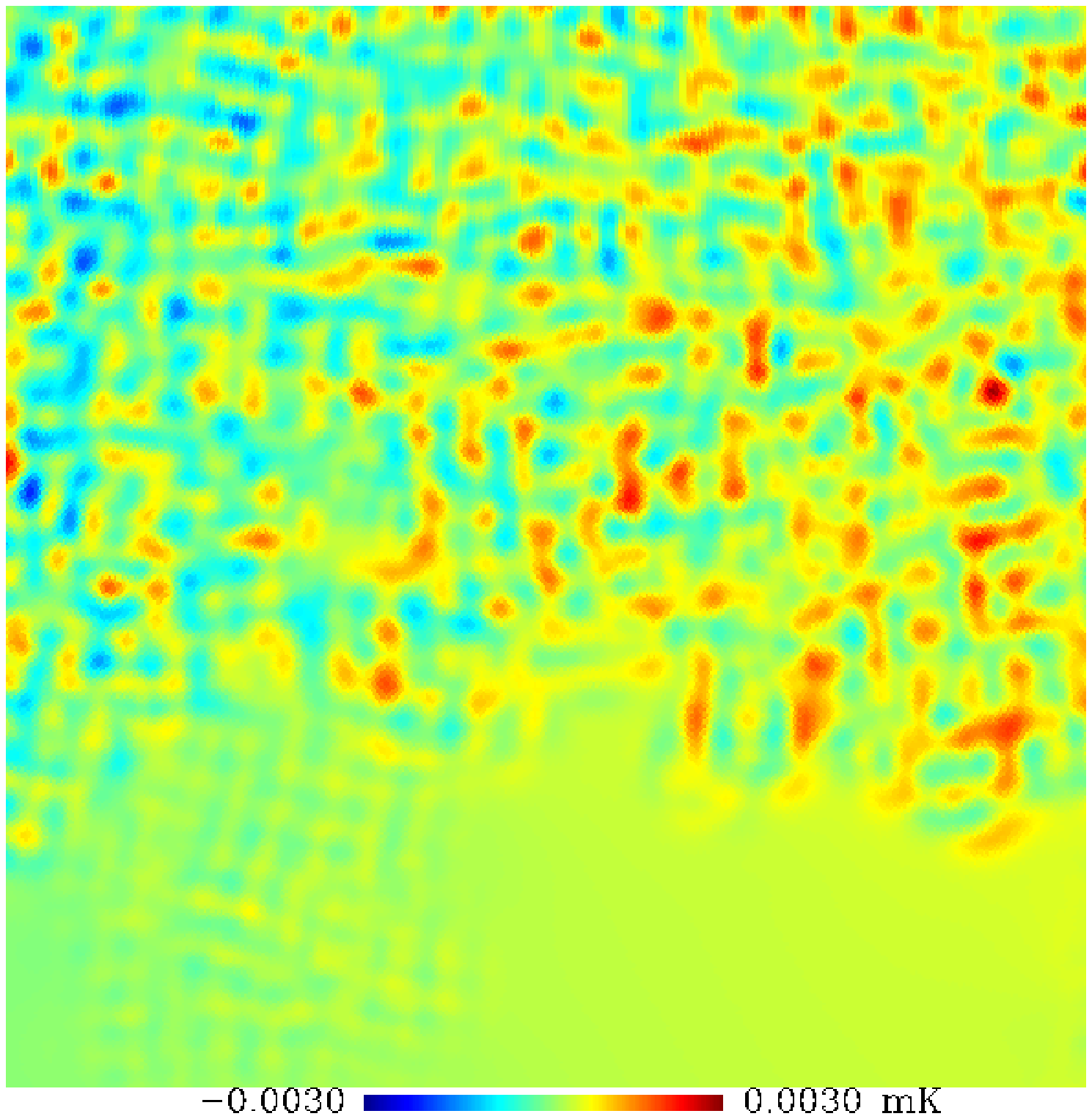}{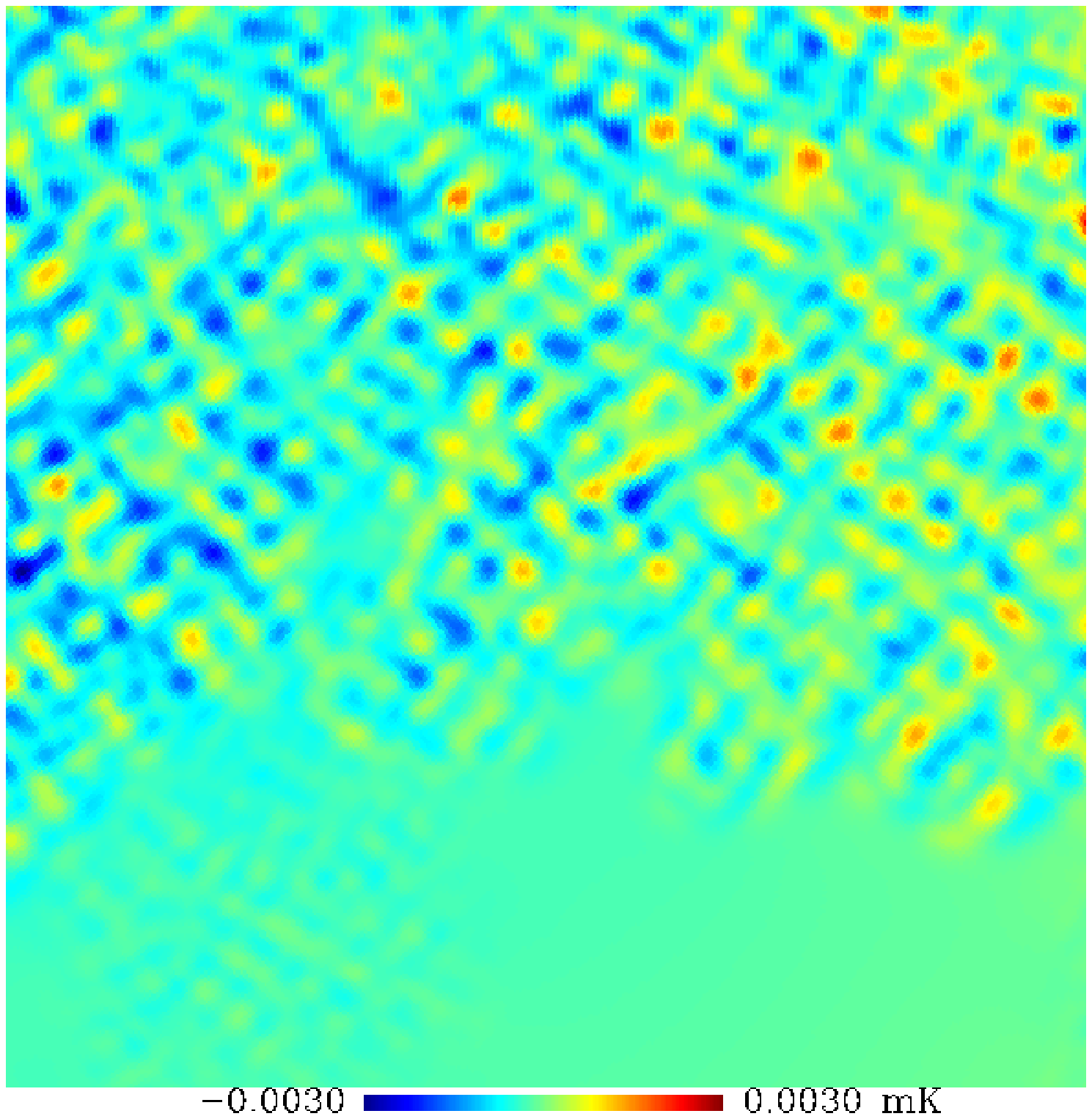}{fig:wiener_pol}%
{Polarization reconstruction. Using the \wmap\ polarization mask
  (\emph{left panel}) for the analysis, we obtain Wiener filtered maps
  for the Stokes parameters Q (\emph{middle panel}), and U
  (\emph{right panel}). The patches are $25 \degr$ on the side,
  centered at galactic coordinates $(l, \, b) = (0 \degr, \, 35
  \degr)$.}

\section{Discussion and conclusions}
\label{sec:summary}

We presented an efficient technique to calculate the Wiener filter
solution of general data sets. Introducing a stochastic auxiliary
field with a simple covariance matrix proportional to the identity
matrix, we indicated how an iterative solver for the Wiener filter
equation can be constructed. The resulting algorithm is not only
numerically robust, but also flexible. Using CMB temperature and
polarization data obtained by the \wmap\ mission as an example, we
explicitly demonstrated the usefulness of the method.

We estimated the numerical efficiency of the algorithm for typical CMB
temperature maps as obtained with the high frequency instrument aboard
the \planck\ satellite \citep{2011A&A...536A...1P}. At Healpix
resolution $\nside = 2048$ and $\lmax = 3000$, calculating the Wiener
filter solution takes about 22 CPUhrs on a single computing node
comprising two quad-core Intel Xeon 2.7~GHz processors. We conclude,
that the method described here is not only applicable to a wide range
of problems, but also effective enough to be of general interest. We
are currently exploring several further variants of the general
messenger approach as well as generalizations of this method to the
case of multiple data sets measuring the same underlying signal with
different transfer functions or more general noise covariance
matrices.

\begin{acknowledgements}
  We thank the anonymous referee for comments which helped to
  improve the presentation of our results. We are grateful to
  S. Prunet for comments on the draft of this paper. BDW was supported
  by the ANR Chaire d'Excellence and NSF grants AST 07-08849 and AST
  09-08902 during this work. FE gratefully acknowledges funding by the
  CNRS. Some of the results in this paper have been derived using the
  HEALPix \citep{2005ApJ...622..759G} package.
\end{acknowledgements}

\bibliographystyle{aa}
\bibliography{literature}

\end{document}